# Influence of the AlN interlayer thickness on the photovoltaic properties of In-rich AlInN on Si heterojunctions deposited by RF sputtering

S. Valdueza-Felip[1*], A. Núñez-Cascajero[1], R. Blasco[1], D. Montero[2], L. Grenet[3], M. de la Mata[4], S. Fernández[5], L. Rodríguez-De Marcos[6], S. I. Molina[4], J. Olea[2], F. B. Naranjo[1]

[1] *Photonics Engineering Group, University of Alcalá, 28871 Alcalá de Henares, Spain*
[2] *Applied Physics Dept. III, University Complutense of Madrid, 28040 Madrid, Spain*
[3] *CEA-Grenoble, LITEN, and University Grenoble Alpes, 38000 Grenoble, France*
[4] *Institute of Electron Microscopy and Materials, Dept. of Materials Science and I. M. and Q. I, Faculty of Sciences, University of Cádiz, 11510, Puerto Real, Spain*
[5] *Research Centre for Energy, Environment and Technology, 28040 Madrid, Spain*
[6] *Thin Film Optics Group, Optics Institute - CSIC, 28006 Madrid, Spain*

## Abstract

We report the influence of the AlN interlayer thickness (0-15 nm) on the photovoltaic properties of $Al_{0.37}In_{0.63}N$ on Si heterojunction solar cells deposited by radio frequency sputtering. The poor junction band alignment and the presence of a 2-3 nm thick amorphous layer at the interface mitigates the response in devices fabricated by direct deposition of $n$-AlInN on $p$-Si(111). Adding a 4-nm-thick AlN buffer layer improves the AlInN crystalline quality and the interface alignment leading to devices with a conversion efficiency of 1.5% under 1-sun AM1.5G illumination. For thicker buffers the performance lessens due to inefficient tunnel transport through the AlN. These results demonstrate the feasibility of using In-rich AlInN alloys deposited by radio frequency sputtering as novel electron-selective contacts to Si-heterojunction solar cells.

---

[*] sirona.valduezafelip@uah.es

III-nitride semiconductors are promising materials for novel electron-selective contacts to *p*-type silicon heterojunction solar cells [1] due to the continuous tunability of their direct bandgap within the solar spectrum from the ultraviolet (AlN: 6.2 eV) to the near infrared (InN: 0.7 eV) using AlInN and InGaN alloys [2]. Their physical and chemical stability enable them to operate in harsh environments, they show superior resistance to high-energy particle radiation and high thermal stability under solar concentration. Additionally, they are environment-friendly from the production viewpoint.

Structures based on InGaN and InN are responsible for visible to near infrared devices, including InGaN-based photovoltaics [3]. The achievement of high-quality InN has been difficult due to its low dissociation temperature (~500ºC under vacuum) compared to AlN and GaN, decreasing the optimum growth temperature for InN. Radio-frequency (RF) sputtering offers the possibility to deposit InN at low temperature with a reasonable crystalline quality, although it presents higher residual carrier concentration (up to $10^{21}$ cm$^{-3}$) [4]. Consequently, bandgap energies around 1.6 eV are found due to the Burstein-Moss effect [5]. RF sputtering is a low-cost technique for the synthesis of large-area nanocrystalline III-nitrides on crystalline and amorphous substrates easy to be scaled to industry. Moreover, it is a simpler growth technique compared to plasma-enhanced chemical vapor deposition, usually used for standard amorphous/crystalline Si heterojunction devices.

In-rich AlInN alloys deposited by RF sputtering are excellent candidates to integrate Si photovoltaics. Liu *et al*. demonstrated the first solar cell deposited by RF magnetron sputtering with a single junction of *n*-Al$_{0.27}$In$_{0.73}$N/*p*-Si(001), a bandgap energy at 2.1 eV and a substrate temperature of 600ºC. An open circuit voltage $V_{oc}$ = 0.25 V, short circuit current density $J_{sc}$ = 30 mA/cm$^2$ and fill factor FF = 29% were achieved under 1-sun AM1.5G illumination, leading to a conversion efficiency of 1.1% [6]. They claimed that this value was limited by the series resistance due to interfacial defects and/or a bad band alignment.

The successful fabrication of high-quality AlInN on Si heterojunctions requires proper control of some issues to reduce shunt pathways and recombination centers, such as the improvement of the alloy crystalline quality and the reduction of carrier recombination at the AlInN/Si interface. Polarization effects are also critical in III-nitrides since can produce band bending and barriers, thus degrading the collection efficiency.

We have previous experience on the growth In-rich $Al_xIn_{1-x}N$ ($x = 0-0.39$) on sapphire [7] and Si(111) [8], with a bandgap energy of 1.7−2.1 eV. However, the lack of substrates for homoepitaxial growth forces the introduction of a buffer to improve the quality of the nitride film on Si. This buffer accommodates the lattice mismatch between both films reducing the accumulated stress and helping the nucleation. Different layers have been used as buffer to improve the crystalline quality of InN and AlInN, being the most frequently used low-growth rate InN [9], GaN and AlN [10,11]. Here we address the effect of inserting an AlN interlayer (thickness of 0-15 nm) on the material quality and electrical properties of $n$-$Al_{0.37}In_{0.63}N$ on $p$-Si(111) heterojunctions deposited by RF magnetron sputtering.

$Al_xIn_{1-x}N$/(AlN) samples were deposited on $p$-doped ($1.5\times10^{14}$–$1.5\times10^{16}$ cm$^{-3}$) 500 µm thick Si(111) with a resistivity of 10-100 Ω·cm using a reactive RF sputtering system, 2'' confocal magnetron cathodes of pure In (4N5) and pure Al (5N), and pure nitrogen (6N) as reactive gas. Substrates were chemically cleaned in organic solvents before being loaded in the chamber where they were outgassed for 30 min at 550°C. Afterwards, substrates were cooled down to the growth temperature. Prior deposition, targets and substrate were cleaned using a soft plasma etching with Ar (99.999%) in the growth chamber. Optimized AlN and $Al_xIn_{1-x}N$ layers were deposited with a nitrogen flow of 6 sccm at a pressure of 0.47 Pa. The substrate temperature was fixed to 450ºC to obtain compact layers [8]. The RF power applied to the Al target was 225 W for the AlN and 150 W for the AlInN, and the RF power applied to the In target was fixed at 40 W. With these growth conditions, the AlInN layer present an Al mole fraction of $x$~0.37, an absorption band edge at 2.1 eV, and a carrier concentration of $n$ ~$2.7\times10^{20}$ $cm^{-3}$ [8]. The AlInN is 80 nm thick, while the insulating AlN buffer thickness ($d$) was varied nominally from 0 to 15 nm (see Table I). Samples under study are $Al_{0.37}In_{0.63}N$/(AlN)/Si structures with $d$ = 0 nm (S1), 4 nm (S2), 8 nm (S3), and 15 nm (S4). The morphological and surface morphology characterization of the samples is detailed in Ref. [12].

The band diagram simulation of these structures was performed using the nextnano[3] software [13], taking into account both spontaneous and piezoelectric polarization, and assuming metal polarity. AlN bandgap energy and Si doping were fixed to 6.25 eV and $5.0\times10^{15}$ $cm^{-3}$, respectively. The *pn* junction shows a built-in voltage of $qV_{bi}$ ~ 1.0 eV.

The films crystalline orientation, composition and mosaicity were estimated from high-resolution x-ray diffraction (HRXRD) data using a PANalytical X'Pert Pro MRD diffractometer. The microstructure of the samples was studied with high-resolution transmission electron microscopy (TEM) techniques performed in a Jeol2100 TEM microscope equipped with a $LaB_6$ gun operated at 200 kV. The optical reflectivity was measured at normal incidence using a Perkin-Elmer Lambda 950 spectrophotometer in the 375-1100 nm range.

Samples were processed into devices with $1\times1$ $mm^2$ mesas defined by Ar etching ($P_{RF}$ = 40 W) in the sputtering system at a pressure of 5 mTorr. The *n*-AlInN contact consists of e-beam evaporated Ti/Al/Ni/Au (30/70/20/100 nm) fingers spaced 150 µm. The *p*-Si(111) contact surrounding the mesas consists of e-beam evaporated Ti/Al (50/100 nm) annealed for 5 min in nitrogen atmosphere at 550°C [14]. *N*- and *p*-contact resistivities of 5 mΩ·cm$^2$ and 6 Ω·cm$^2$ were obtained from transmission line method measurements. A schematic description of the AlInN/AlN/Si solar cell structure with a top-view optical microscopy image of a device with 1×1 mm$^2$ mesa is presented in Fig. 1(a).

Dark current density-voltage (J-V) measurements and under 1 sun AM1.5G solar illumination were recorded using a semiconductor parameter analyzer. The spectral response was measured with a 250 W halogen lamp coupled to a monochromator and calibrated with a reference Si photodetector. A 405-nm laser with an output power of 0.7 µW was used to precisely determine the external quantum efficiency (EQE) of the cells at that wavelength. For voltage-dependent EQE measurements, devices were connected to a bias source in series with a load resistor smaller than the device resistance. Data were extracted measuring the voltage drop across the load resistor vs the bias.

The structural quality of the AlInN films as a function of the AlN buffer thickness was evaluated with HRXRD measurements. Fig. 1(b) shows the diffractograms corresponding to the 2θ/ω scans of the layers. Only the peaks related to the Si(111) and the AlInN (0002) and (0004) diffraction are observed, pointing out a wurtzite structure highly oriented along the *c*-axis for the nitride films, with the absence of the typical multiple peaks measured in layers with phase separation. The increase of the intensity of the (0004) AlInN diffraction peak with the AlN thickness in samples S2-S4 is a sign of improvement of the AlInN crystal quality when incorporating the buffer. The diffraction peak related to the (0002) reflection of AlN can be appreciated in the sample with a 15-nm thick buffer. The FWHM of the AlInN rocking curve decreases from 8º for S1 (without buffer) to 5º when introducing an AlN buffer of 4–15 nm thickness, confirming the observed AlInN structural quality gain.

High-resolution TEM measurements, performed in similar samples grown under the same growth conditions, were carried out in order to visualize the interface between the nitride layer and the Si substrate. Figure 2 (a) displays a high-resolution TEM image of the Si/AlInN interface of sample S1 without AlN buffer showing a rough thin amorphous layer (2-3 nm thick), which has been assigned to native silicon oxide or silicon nitride (unintentionally created by surface nitridation of the Si substrate). Magnified details of both crystalline phases involved, AlInN and Si, are included along with their indexed FFTs in Fig. 2(b) and (c). It is worth to note the clear epitaxial relationship between the Si and AlInN despite the presence of the amorphous interlayer, which can be described as follows: $(0001)[11\text{-}20]_{AlInN}\|(11\text{-}1)[112]_{Si}$. This interlayer at the AlInN/Si interface is unintentionally present at the Si surface in the case of silicon oxide or formed at the initial stage of the nitride growth in the case of silicon nitride. This amorphous layer is not observed in AlInN/Si samples containing an AlN buffer [12].

Fig. 3(a) shows the J-V characteristics of the fabricated devices with structures S1-S4 under dark conditions showing a rectifying behavior. The saturation current density ($J_0$), series and shunt resistances ($R_s$, $R_{sh}$) and the diode ideality factor ($\eta$) were estimated fitting experimental data to the expression: $J = J_0 \cdot [\exp(V_d/(\eta \cdot V_T))\text{-}1]+V_d/R_{sh}$, where $V_d$ is the diode voltage ($V_d = V\text{-}J \cdot R_s$), and $V_T$ is the thermal voltage given by $V_T = k_B \cdot T/q$ (26 meV at room temperature). From the results summarized in Table I, we observe that $R_s$ increases with the AlN thickness, as expected from its insulating behavior, with the exception of S2 for which it slightly decreases. $R_{sh}$ also increases with the AlN thickness, a sign of improved structural quality of the junction as previously discussed from HRXRD data. At the same time, the reverse saturation current $J_0$ does not show any clear dependence with the buffer thickness, pointing to undesired carrier recombination as the main component of the measured leakage current.

J-V measurements under 1 sun AM1.5G illumination [Fig. 3(b)] were carried out to evaluate the photovoltaic characteristics of the devices, which values are summarized in Table I. Comparing the four devices, the $V_{oc}$ and the $J_{sc}$ decrease respectively from 270 to 80 mV and from 15.60 to 0.47 mA/cm$^2$ when thickening the AlN buffer from 0 to 15 nm, whereas the FF remains ~16%. This performance deterioration occurs since the transport of the carriers via tunneling through the AlN buffer is hindered so they are not efficiently collected at the contacts [15].

The quality gain of the AlInN layer with 4-nm thick AlN buffer is also noticed in the results from J-V curves under illumination, pointing to a maximum $V_{oc}$ = 350 mV, $J_{sc}$ = 22.2 mA/cm$^2$, FF = 20% and conversion efficiency of 1.5% without any antireflection layer and/or back reflector. These data represent an improvement compared with results reported in 90-nm-thick $n$-Al$_{0.27}$In$_{0.73}$N/$p$-Si(001) junctions deposited without buffer layer by RF sputtering obtaining 1.1% [6]; and in $n$-Al$_{0.54}$In$_{0.46}$N/AlN/$p$-Si(001) devices with 170 nm of AlInN and similar AlN thickness deposited at 700ºC by sputtering, obtaining $V_{oc}$ = 0.3 V and $J_{sc}$ = 9.1 µm/cm$^2$ [10]. It has to be noted that in these structures the AlN and AlInN layers were respectively deposited in RF and DC mode, and own higher Al content (46-57%).

These low values of $V_{oc}$ (< 350 mV), $J_{sc}$ (< 23 mA/cm$^2$) and FF (< 20%) occur due to the pronounced S-shape of the illuminated J-V curves. Other research groups have reported similar S-shape features in AlInN/Si junctions [6], InGaN/GaN multiple-quantum well solar cells [15] and MoO$_x$/$n$-type c-Si devices [16] among others. This effect indicates a reduction of the generated photocurrent due to inefficient carrier collection pointing to the existence of a barrier for electrons or a highly recombinative AlInN/Si interface.

The EQE represents the fraction of collected electron-hole pairs per incident photon and was calculated in the 375-1100 nm range as EQE = ($J_{op}$/$P_{op}$)·(hc/qλ), where $J_{op}$ is the photocurrent density, $P_{op}$ the optical power density impinging the device, q the electron charge, h is Planck's constant, c is the speed of light, and λ the wavelength of the incident light. The internal quantum efficiency (IQE) was deduced from reflectivity spectra, R(λ), as IQE(λ) = EQE(λ) / [1-R(λ)], which represents the fraction of collected electron-hole pairs per absorbed photon.

The EQE and IQE curves of the devices are displayed in Fig. 4(a). The EQE presents a peak in the visible spectral range above the AlInN bandgap energy followed by a flat photoresponse in the near infrared range with a cutoff around 1100 nm related to the Si band edge. EQE and IQE values at 560 nm are summarized in Table I. For devices without buffer (S1), the peak EQE is reduced (27% at 490 nm) due to the poor band alignment of the conduction and valence band edges at the heterointerface as displayed in the band diagram simulations of Fig. 4(b).

Best results in terms of EQE and IQE are achieved for sample S2 with $d$ = 4 nm, which shows the highest EQE = 44% and IQE = 53% at 560 nm. This improvement can be due to (i) the improved structural quality of the AlInN layer, and (ii) the higher separation energy between the conduction and valence band edges at the Si interface thanks to the introduction of the AlN, that compensates the polarization-induced charges at the interface and thus the internal electric field. As depicted in Fig. 4(b), in a pure AlInN/Si interface, these internal fields would drift the photogenerated carriers towards the heterojunction, thus reducing the carrier collection through the contacts. In the case of sample S2, the AlN buffer is thin enough to permit the tunnel transport through it [17]. However, for thicker buffers the number of collected carriers is reduced due to inefficient transport through the AlN barrier for both electrons and holes, in agreement with the band profile of the samples shown in the inset of Fig. 4(b). In fact, the EQE drops at 560 nm to a value of 5% and 0.10% for S3 and S4, respectively.

To investigate the origin of the S-shape in the illuminated J-V curves, EQE measurements of sample S2 ($d = 4$ nm) were taken at various bias voltages from -1.5 V (i.e. negative pole to the *p*-side) to +0.15 V (i.e. positive pole to *p*-side): below and above the knee voltage $V_K \sim -0.10$ V. As shown in Fig. 5, the zero-bias EQE spectrum shows a reduced spectral response beyond the AlInN band edge at 590 nm, pointing out that carriers generated in the Si region are not fully collected. As the device is reverse biased, the device photoresponse increases in the range between 590 nm (AlInN band edge) and 1100 nm (Si band edge), indicating that the generation and collection of photocarriers is empowered and the depletion width broadens towards Si. From nextnano$^3$ simulations, we estimate that the depletion width covers ~480 nm of Si for $V_{in} = 0$ V, and it extends towards ~630 nm when applying an external bias of $V_{in} = -0.6$ V. The contribution of the AlInN to the depletion width is almost negligible. Inset of Fig. 5 depicts the effect of the bias voltage on the EQE below the AlInN bandgap energy at 700 nm, achieving 64% at 700 nm when reverse biasing at -1.5 V. We note that a saturation effect is reached at $V_{in} \sim -0.6$ V, approximately in accordance with the end of the kink of the J-V curve in Fig. 2(b).

In summary, photovoltaic devices based on *n*-Al$_{0.37}$In$_{0.63}$N/AlN/*p*-Si(111) heterojunctions were fabricated by RF sputtering varying the AlN interlayer thickness (0-15 nm). Best results were obtained for devices with a 4 nm of AlN, namely $V_{oc} = 350$ mV, $J_{sc} = 22.2$ mA/cm$^2$, and FF = 20% under 1 sun AM1.5G illumination, leading to a conversion efficiency of 1.5%. The spectral response covers the 375-1100 nm range with an $EQE_{peak}$ of 44% at 560 nm, and increases to 64% at 700 nm when reverse biasing at -1.5 V. For thicker buffers the photovoltaic performance drops due to inefficient carrier tunnel transport through the buffer.

**Acknowledgments**

We thank E. Monroy from the CEA-Grenoble and University Grenoble Alpes for useful discussions; and J. L. Thomassin from the PTA-CEA-Grenoble and J. A. Méndez from the CSIC-Madrid, for technical support. Support from projects NitPho (TEC2014-60483-R), ANOMALOS (TEC2015-71127-C2-2-R), INFRASIL (TEC 2013-41730-R), SINFOTON (S2013/MIT 2790), MADRID-PV (2013/MAE-2780), PhotoAl (CCG2015/EXP-014), PAI research group (TEP-946 INNANOMAT), and FEDER-EU is acknowledged. TEM data were taken at DME-SC-ICyT-UCA. A. Núñez-Cascajero thanks her grant to the University of Alcalá and D. Montero acknowledges his contract BES-2014-067585.

**Table I**

Electrical characteristics of the $Al_{0.37}In_{0.63}N$ on Si heterojunctions vs the AlN buffer thickness.

| Sample | AlN thickness (nm) | $R_s$ ($\Omega \cdot cm^2$) | $R_{sh}$ ($k\Omega \cdot cm^2$) | $J_0$ ($\mu A/cm^2$) | $\eta$ | $V_{oc}$ (mV) | $J_{sc}$ ($mA/cm^2$) | FF (%) | Eff. (%) | EQE at 560 nm (%) | IQE at 560 nm (%) |
|---|---|---|---|---|---|---|---|---|---|---|---|
| **S1** | 0 | 1.9 | 200 | 8.0 | 6.0 | 270 | 15.60 | 16 | 0.7 | 21.90 | 25.80 |
| **S2** | 4 | 0.9 | 300 | 5.0 | 4.5 | 350 | 22.20 | 20 | 1.5 | 43.60 | 53.40 |
| **S3** | 8 | 4.8 | 300 | 8.0 | 6.5 | 240 | 9.60 | 17 | 0.4 | 4.80 | 7.50 |
| **S4** | 15 | 38.5 | 500 | 7.0 | 8.2 | 80 | 0.47 | 16 | $6.0 \times 10^{-3}$ | 0.10 | 0.14 |

**Figure captions**

Fig. 1. (a) Schematic description of the AlInN/AlN/Si solar cell structure. Inset: Top-view optical microscopy image of a device with 1×1 mm$^2$ mesa. (b) HRXRD 2θ/ω scans of samples S1-S4.

Fig. 2. (a) HRTEM image of the Si/AlInN interface evidencing the presence of an intermediate amorphous layer between both materials. Images (b) and (c) display magnified details of the AlInN and Si respectively, along with their indexed FFTs.

Fig. 3. Current density vs voltage curves of devices S1-S4 measured (a) in the dark including the fitted curves (dashed lines), and (b) under 1 sun of AM1.5G illumination. The knee voltage is defined as $V_K$ ~ -0.20, -0.10, -0.30, and -0.95 V for samples S1 to S4, respectively. Inset: J-V curve in the 4$^{th}$ quadrant.

Fig. 4. (a) EQE (solid lines) and IQE (dashed lines) of devices S1-S4 measured at $V_{in}$ = 0 V. The AlInN bandgap wavelength is marked with a dashed line. (b) Energy band diagram of the *p*-Si(111)/(AlN)/*n*-Al$_x$InN structures with 0, 4, 8 and 15 nm of AlN (samples S1-S4) and $V_{in}$ = 0 V. Inset: detail of the band diagram.

Fig. 5. EQE of device S2 biased with $V_{in}$ = (-0.15-1.5) V. Inset: sketch of the bias-dependent EQE measurement setup. Inset: evolution of the EQE at 700 nm vs the input bias for S2.

Figure 1

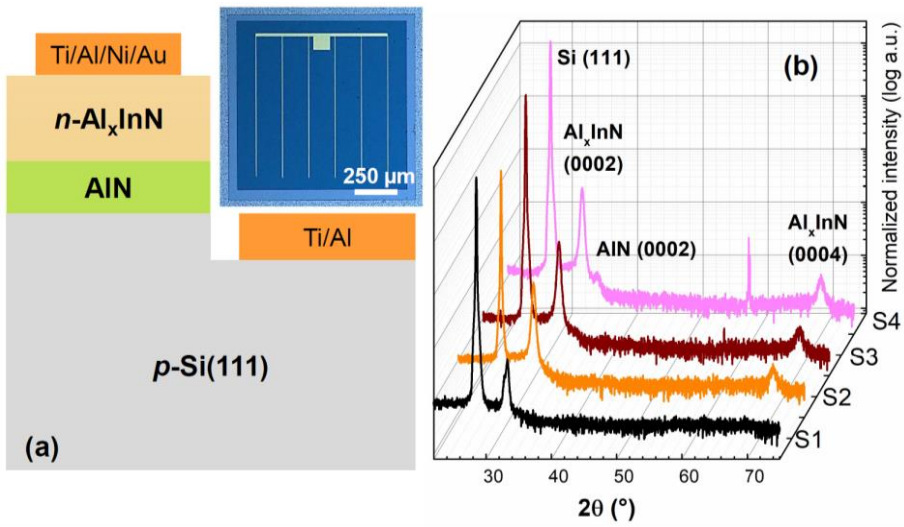

Figure 2

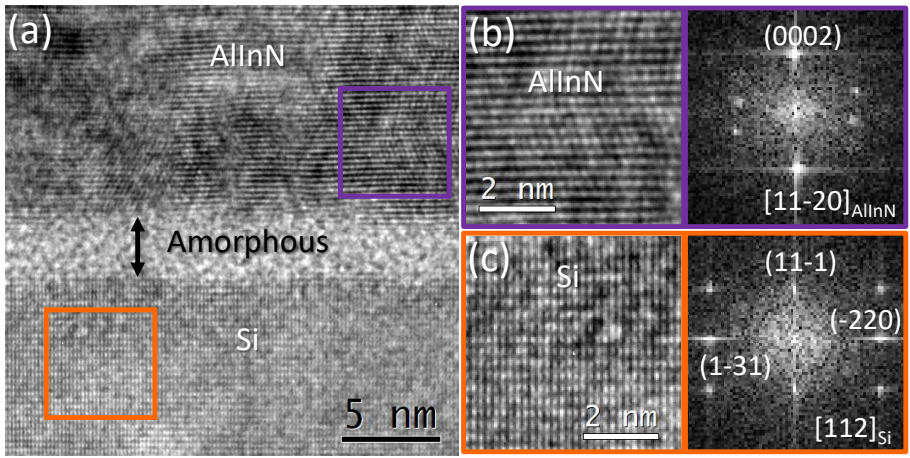

Figure 3

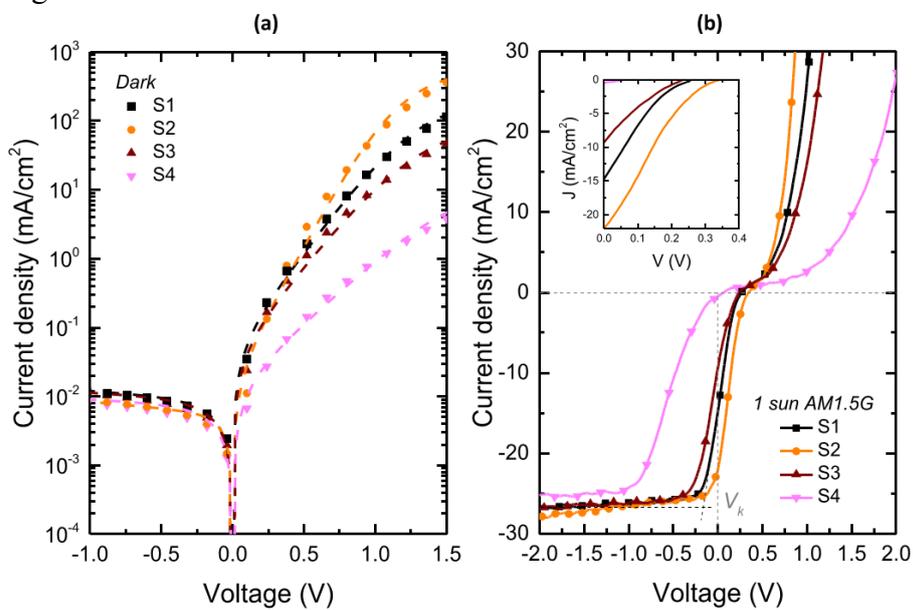

Figure 4

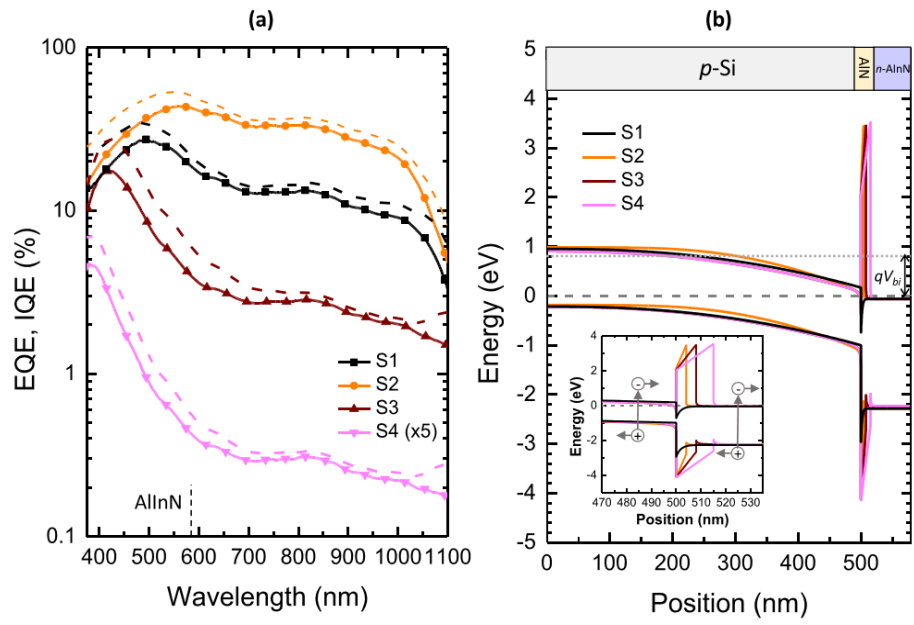

Figure 5

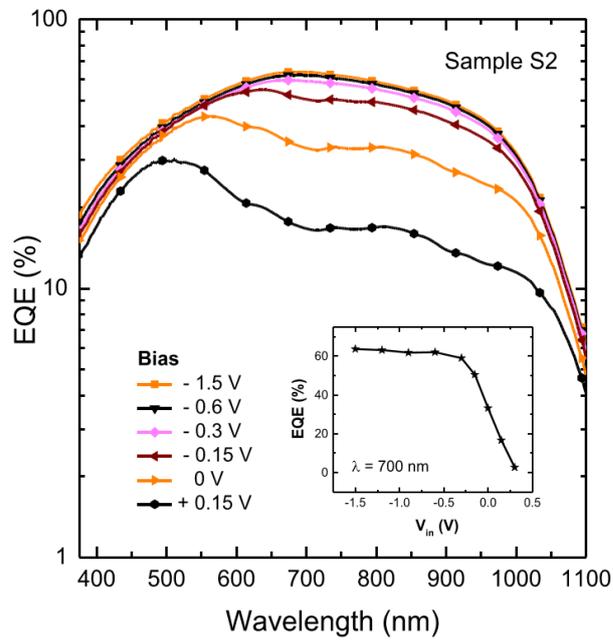